\documentclass[10pt,conference]{IEEEtran}

\usepackage{microtype}
\usepackage{xspace}
\usepackage{cite}
\usepackage{balance}
\usepackage{amsfonts,amsmath,amsthm}
\usepackage{amsthm}
\usepackage{graphicx}
\usepackage{hyperref}
\usepackage[utf8]{inputenc}
\usepackage{algorithm}
\usepackage[noend]{algpseudocode}
\usepackage{epigraph}
\usepackage[T1]{fontenc}
\usepackage{color}
\usepackage[utf8]{inputenc}
\usepackage[english]{babel}
\usepackage{amsthm}
\usepackage{tikz}
\usepackage{standalone} 

\theoremstyle{definition}

\definecolor{blueLink}{rgb}{0,0.2,0.8}
\hypersetup{colorlinks,linkcolor=blueLink,urlcolor=blueLink,citecolor=blueLink}

\newcommand{\lref}[2]{\hyperref[#2]{#1~\ref*{#2}}}

\theoremstyle{plain}
\newtheorem{theorem}{Theorem}
\newtheorem{lemma}{Lemma}

\newcommand{\OPT}{\textsc{Opt}\xspace}
\newcommand{\ALG}{\textsc{Alg}\xspace}
\newcommand{\Nat}{\mathbb{N}}
\newcommand{\Inp}{\mathcal{I}}

\newcommand{\linkSet}{E_{\ell}}
\newcommand{\Packets}{\Pi}
\newcommand{\PacketsLinks}{\Packets_{\ell}}
\newcommand{\dist}[1]{d{\left(#1\right)}}
\newcommand{\Dist}[1]{\hat{d}(#1)}
\newcommand{\fixedDist}[1]{d_{\ell}(#1)}
\newcommand{\size}[1]{\textsf{size}(#1)}
\newcommand{\source}[1]{\textsf{src}(#1)}
\newcommand{\dest}[1]{\textsf{dest}(#1)}

\newcommand{\before}[1]{B_{#1}} 
\newcommand{\heavier}[2]{H_{#1}(#2)}
\newcommand{\lighter}[2]{L_{#1}(#2)}
\newcommand{\Heavier}[1]{H_{#1}}
\newcommand{\Lighter}[1]{L_{#1}}
\newcommand{\adj}[2]{A_{#1}(#2)}
\newcommand{\impE}[2]{\Delta_{#1}(#2)}

\newcommand{\chunks}[2]{C_{#1}(#2)}
\newcommand{\match}[1]{M_{#1}}

\newcommand{\weight}[1]{w_{#1}}
\newcommand{\Weight}[1]{w{\left({#1}\right)}}
\newcommand{\edge}[1]{e_{#1}}
\newcommand{\Edges}[1]{E_{#1}}
\newcommand{\release}[1]{a_{#1}}
\newcommand{\trans}{T}
\newcommand{\receiver}{R}
\newcommand{\dual}{\ensuremath{\mathcal{D}}\xspace}
\newcommand{\primal}{\ensuremath{\mathcal{P}}\xspace}

\newcommand{\act}[1]{A(#1)}

\title{Scheduling Opportunistic Links in\\Two-Tiered Reconfigurable Datacenters}

\author{\IEEEauthorblockN{Janardhan Kulkarni}
\IEEEauthorblockA{
Microsoft Research, Redmond, USA \\
jakul@microsoft.com}
\and
\IEEEauthorblockN{Stefan Schmid} \IEEEauthorblockA{
University of Vienna, Austria \\
stefan\_schmid@univie.ac.at}
\and
\IEEEauthorblockN{Pawe{\l} Schmidt} \IEEEauthorblockA{University of Wroc{\l}aw, Poland\\
pawel.schmidt@cs.uni.wroc.pl}
\thanks{Research supported by the
European Research Council (ERC),
consolidator grant agreement 864228
(AdjustNet) and by Polish National Science Centre grant 2016/22/E/ST6/00499.}
}

\date{}
\IEEEoverridecommandlockouts 
\begin{document}

\maketitle

\begin{abstract}
Reconfigurable optical topologies are emerging as a promising technology to
improve the efficiency of datacenter networks. This paper considers the problem
of scheduling opportunistic links in such reconfigurable datacenters. We study
the online setting and aim to minimize flow completion times. The problem is a
two-tier generalization of classic switch scheduling problems. We present a
stable-matching algorithm which is $2\cdot (2/\varepsilon+1)$-competitive
against an optimal offline algorithm, in a resource augmentation model: the
online algorithm runs $2+\varepsilon$ times faster. Our algorithm and result are
fairly general and allow for different link delays and also apply to hybrid
topologies which combine fixed and reconfigurable links. Our analysis is based
on LP relaxation and dual fitting.
\end{abstract}

\section{Introduction}

Given the popularity of data-centric applications and machine learning, the
traffic in datacenters is growing explosively. Accordingly, over the last years,
great efforts were made to render these networks more efficient, on various
layers of the networking stack~\cite{singh2015jupiter}, including the physical
network topology, e.g.,~\cite{clos,bcube,slimfly,singla2012jellyfish,xpander}.  
As a next frontier toward more efficient datacenter networks reconfigurable
optical topologies are emerging
\cite{opera,rotornet,helios,cthrough,projector,avin2017demand,ballani2020sirius},
and in particular, demand-aware topologies such as
\cite{helios,cthrough,projector,avin2017demand,splaynet,eclipse,xweaver} which
can dynamically adapt towards the traffic patterns they serve. This is
attractive as empirical studies show that datacenter workloads are indeed skewed
and bursty, featuring much temporal and spatial structure
\cite{facebook,benson2010network,tracecomplexity}, which may be exploited in
adaptive infrastructures. For example, these technologies allow to flexibly
transmit elephant flows via opportunistic links that provide shortcuts between
the frequently communicating racks.

A key issue for the efficient operation of reconfigurable datacenter networks
concerns the scheduling of the opportunistic links. As the number of these links
is limited, they should be used for the most significant transmissions. This
however is challenging as scheduling decisions need to be performed in an online
manner, when the demand is not perfectly known ahead of time.

This paper studies this scheduling problem from a competitive analysis
perspective: we aim to design an online scheduling algorithm which does not
require any knowledge about future demands, but performs close to an optimal
offline algorithm which knows the entire demand ahead of time. In particular, we
consider a two-stage switch scheduling model as it arises in existing datacenter
architectures, e.g., based on free-space optics~\cite{projector}. In a nutshell
(a formal model will follow shortly), we consider an architecture where traffic
demands (modelled as \emph{packets}) arise between Top-of-Rack (ToR) switches,
while opportunistic links are between lasers and photodetectors, and where many
laser-photodetector combinations can serve traffic between a pair of ToRs. The
goal is to minimize the packet (i.e., flow) completion times over all packets in
the system. 

The problem is reminiscent of problems in classic switch scheduling
\cite{mckeown1999islip,chuang1999matching}, as in each time step, an optical
switch allows to ``transmit a matching''. However, the two-stage version turns
out to introduce several additional challenges (as also pointed out in
\cite{projector}), which we tackle in this paper.

\subsection{Our Contribution}

This paper initiates the study of online switch scheduling algorithms for a
multi-stage model which is motivated by emerging reconfigurable optical
datacenter architectures. Our main contribution is an online stable-matching
algorithm that is $2\cdot(2/\varepsilon+1)$-competitive against a powerful
hindsight optimal algorithm, in a resource augmentation model in which the
online algorithm runs $2+\varepsilon$ times faster. Our analysis relies linear
programming relaxation and dual fitting: we formulate primal and dual linear
programs to which we will charge the costs of our online algorithm.

Our algorithm and result allow for different link delays and also apply to
hybrid datacenter topologies as they are often considered in the literature:
topologies which combine fixed and reconfigurable links. 

We emphasize that resource augmentation is necessary for obtaining competitive
algorithms: Dinitz et al. \cite{dinitz2020scheduling} prove that even in single
tier networks, no randomized algorithm can be competitive against an adversary with
matching transmission speed.

\subsection{Overview of the Algorithm and Techniques: Analysis via Dual-Fitting}

Our algorithm for the problem is based on a generalization of stable matching
algorithm 
\cite{gale1962college}
for the two-tier networks. Informally, 
in our algorithm, each transmitter maintains a queue
of packets that are not scheduled yet. The packets in the queue are sorted in
the decreasing order of weights. At each time step, our algorithm finds a stable
matching between transmitters and receivers as follows: In our
problem we are given a bipartite graph $B = \{ (T \cup R), E \}$, between the set of
transmitters $T$ and set of receivers $R$; the edge set $E$ denotes the
connections between transmitters and receivers. At time step $\tau$, we assign
the edge~$e=(t, r)$ connecting the transmitter $t$ to the receiver~$r$ a weight~$w_e$, which is equal the highest weight packet in the queue of transmitter $t$
at time instant $\tau$ which wants to use the edge~$e$. Taking the weights of
edges as {\em priorities}, our algorithm simply computes a stable matching $M$
in the graph $B$, and schedules $M$ at time step $\tau$. (Note that since
priorities in our algorithm are symmetric, one can compute a stable matching by
a simple greedy algorithm.)

However, how should we assign an incoming packet to a (transmitter, receiver)
pair? In our algorithm, as soon as a packet arrives it is dispatched to a
specific (transmitter, receiver) pair via which our algorithm commits to
eventually transmitting the packet. This is the decision that complicates
routing in two-tier networks. Our dispatch policy estimates the worst case
impact of transmitting a packet via a specific (transmitter, receiver) pair,
taking into account the set of queued packets in the system. In particular, we
estimate how much latency of the system {\em increases} if a packet is
transmitted via a (transmitter, receiver) pair. Finally, we chose the
(transmitter, receiver) pair which has the least impact. We show that this
greedy-dispatch policy coupled with stable matching is indeed competitive in the speed augmentation model \cite{kalyanasundaram2000speed},
where we assume that the online algorithm can transmit the packets at twice the rate compared to the optimal offline algorithm.
It is not hard to show that without speed augmentation,  no online algorithm can be competitive~\cite{dinitz2020scheduling}.

Our algorithm and its analysis via dual fitting is inspired by scheduling for
unrelated machines~\cite{anand2012resource}, and combines two disparate research
directions: switch scheduling and  scheduling for unrelated machines.

Our analysis via dual-fitting works as follows. First we write a linear
programming relaxation for the underlying optimization problem, and then we take
the dual of the LP. This is done assuming that we know the entire input, which
we can do because LP duality is only used in the analysis. The weak duality
theorem states that {\em any feasible solution} to the dual is a lowerbound on
the optimal primal solution (which in turn is a lowerbound on the optimal
solution to our problem). The crux of the the dual-fitting analysis is to relate
the cost of our algorithm to a feasible dual solution, thus allowing us to
compare our cost to the optimal solution.

The dual of our LP for the problem has a rather interesting form. It consists of
variables $\alpha_p$ for each packet~$p$. For each time step $\tau$, we also
have variables $\beta_{t, \tau}$ for each transmitter $t$ and $\beta_{r, \tau}$
for a receiver $r$. We interpret $\alpha_p$ as the latency seen by packet~$p$
and $\beta_{t, \tau}, \beta_{r, \tau}$ variables as set of packets that are
waiting to use the transmitter $t$ and receiver $r$ at time step $\tau$.
Clearly, sum over $\alpha_p$ variables is equal to the total latency seen by
packets. It is not hard to argue that the same also holds for $\beta$ variables.
However, the crucial part of the analysis is to show that indeed such an
interpretation of the dual variables is a feasible solution. This is done by showing
that our setting of dual variables satisfies all the dual constraints. Verifying
the dual constraints crucially uses both our algorithmic decisions regarding
dispatch policy and the stable matching algorithm. One could also interpret that
our algorithmic decisions were in fact driven by the dual LP, in the sense of
primal-dual algorithms~\cite{buchbinder2009design}. 

\subsection{Organization}

The remainder of this paper is organized as follows. We introduce our formal
model in \lref{Section}{sec:model}. Our algorithm is described in
\lref{Section}{sec:algo} and analyzed in \lref{Section}{sec:analysis}. After
reviewing related work in \lref{Section}{sec:relwork}, we conclude our
contribution in \lref{Section}{sec:conclusion}.

\section{Model}
\label{sec:model}

We consider a hybrid optical network which consists of a fixed and a
reconfigurable topology. We model this network as a graph $G=(V, E, d)$ where
$V$ is the set of vertices partitioned into the following four layers:
sources~$S$, transmitters $T$, receivers $R$, destinations $D$. Each transmitter
$t\in T$ is attached (has an edge) to a particular source $\source{t}$ and each
receiver $r\in R$ is attached to a particular destination $\dest{r}$ (a single
source or destination may have multiple transmitters or receivers attached). The
edges between transmitters and receivers form an optical reconfigurable network.
For transmitter $t\in T$, we denote the set of receivers adjacent to $t$ in $G$
by $R(t)$; and for receiver $r\in R$, $T(r)$ is the set of transmitters adjacent
to $r$ in $G$. The fixed part of the network is a set $\linkSet \subseteq E$ of
direct source-destination \emph{links}. 

At any time $\tau \in \Nat_+$, a transmitter $t$ may have at most one active
edge connecting it with a receiver from $R(t)$, and each receiver $r$ may have
at most one active incoming edge from one of transmitters $T(r)$. For any edge
$e\in E$, the \emph{delay} of that edge is defined by $\dist{e} \in \Nat$, that
is, $s \cdot \dist{e}$ is the time required to transmit a packet of size $s$
through that edge. If $e$ is a transmitter-receiver connection, then its delay
is at least $1$.

\begin{figure*}[t]
    \centering
    \begin{minipage}[c]{0.4\linewidth}
        \tikzstyle{circle}=[fill=gray, draw=black, shape=circle, scale=0.5]
\begin{tikzpicture}
	\node [circle, label={left:$s_1$}] (0) at (-5, 3) {};
	\node [circle, label={left:$s_2$}] (1) at (-5, 1) {};
	\node [circle, label={above:$t_1$}] (2) at (-4, 3.75) {};
	\node [circle, label={above:$t_2$}] (3) at (-4, 2.5) {};
	\node [circle, label={above:$t_3$}] (4) at (-4, 1) {};
	\node [circle, label={above:$r_4$}] (5) at (-2, 1) {};
	\node [circle, label={above:$r_3$}] (6) at (-2, 2) {};
	\node [circle, label={above:$r_2$}] (7) at (-2, 3) {};
	\node [circle, label={above:$r_1$}] (8) at (-2, 3.75) {};
	\node [circle, label={right:$d_1$}] (9) at (-1, 3.75) {};
	\node [circle, label={right:$d_2$}] (10) at (-1, 2.5) {};
	\node [circle, label={right:$d_3$}] (11) at (-1, 1) {};
	\draw (0) to (2);
	\draw (0) to (3);
	\draw (1) to (4);
	\draw (5) to (11);
	\draw (6) to (10);
	\draw (10) to (7);
	\draw (8) to (9);
	\draw [dashed] (2) to (8);
	\draw [dashed] (2) to (7);
	\draw [dashed] (3) to (6);
	\draw [dashed] (4) to (5);
	\draw [dashed] (4) to (6);
	\draw [double, double distance=1, bend right] (1) to (11);
\end{tikzpicture}
    \end{minipage}
    \quad
    \begin{minipage}[c]{0.5\linewidth}
    \begin{center}
    \begin{tabular}{ c | c | c || c | c}
        packet & path & arrival & transmission & edge \\ 
        \hline
        $p_1$ & $s_1 \to d_1$ & $1$ & $1$ & $(t_1,r_1)$ \\
        $p_2$ & $s_1 \to d_2$ & $1$ & $2$ & $(t_1,r_2)$ \\
        $p_3$ & $s_2 \to d_2$ & $1$ & $1$ & $(t_3,r_3)$ \\
        $p_4$ & $s_2 \to d_2$ & $2$ & $2$ & $(t_3,r_3)$ \\  
        $p_5$ & $s_2 \to d_3$ & $2$ & $2$ & $(s_2,d_3)$
    \end{tabular}
    \end{center}
    \end{minipage}
    \caption{Input graph $G$. Solid lines represent edges between sources and
    transmitters or between receivers and destinations. Available connections in
    the reconfigurable layer are shown as dashed lines. The double line
    represents a direct source-destination link in the fixed layer. In this
    example, $\dist{e} = 1$ for any edge in the reconfigurable layer,
    $\dist{s_2, d_3}=4$ and $\dist{e} = 0$ for all other edges. The table
    constains a set of packets (all of unit weight) and a feasible solution. In
    the first transmission step packets $\{ p_1, p_3 \}$ are transmitted, and in
    the second step $\{ p_2, p_4 \}$ are sent through the reconfigurable network
    and $p_5$ via link $(s_2,d_3)$. The cost of this solution is $9$, while the
    cost of the optimal solution of this instance is $7$ (sending $p_5$ in the
    third step via $(t_3,r_4)$).}
    \label{fig:model} 
\end{figure*}

We study the design of a topology scheduler whose input is a sequence of packets
$\Packets$ arriving in an online fashion. A packet $p\in\Packets$ of weight
$\weight{p} > 0$ which arrives at time $\release{p}$ at source node
$\source{p}\in S$, has to be routed to destination $\dest{p} \in D$. For
packet~$p\in\Packets$, let $\Edges{p}$ be the set of transmitter-receiver edges
from the reconfigurable network that might be used to deliver $p$, i.e.,
$\Edges{p} = \{(t,r) \in T\times R : \source{t} = \source{p} \text{ and }
\dest{r} = \dest{p} \}$. If there exists a link connecting $\source{p}$ with
$\dest{p}$, the delay of that link is $\fixedDist{p} = d(\source{p}, \dest{p})$.
By $\PacketsLinks$ we denote the set of all packets that can be transmitted
through the fixed network. 

In this paper, we assume that packets are of uniform size. However, this assumption is without loss of generality in the speed augmentation model.
By standard arguments \cite{anand2012resource}, one can treat a packet~$p$ of size $\ell_p$ as $\ell_p$ unit-length packets 
each of weight $w_p/\ell_p$. Hence, in the rest of the paper we assume that packets of uniform size.

The goal of the algorithm is to route all packets from their sources to
destinations. A packet can be transmitted either through the reconfigurable
network or the slower direct connection (if available) between source and
destination. All transmissions happen only at times $\tau \in \Nat_+$, but
packets may arrive between transmissions. When packet~$p$ arrives at time $\tau
\in (\tau', \tau'+1]$ for some $\tau'\in \Nat_+$, it will be available for
transmission at next transmission slot, namely at time $\tau'+1$. Therefore, we
may assume, that packets arrive only at integral times (i.e., the arrival time
of packet is shifted from $\tau$ to $\lceil \tau \rceil$) and they immediately
can be transmitted through the network. An example input and transmission
schedule are presented on \lref{Figure}{fig:model}.

We define the \emph{weighted fractional latency} as follows: when a fraction $0
< x \leq 1$ of packet~$p$ reaches its destination $\dest{p}$ during transmission
step $\tau$, it incurs the weighted latency of $x \cdot \weight{p} \cdot (\tau +
1 - \release{p})$ (in this case, $\tau+1$ is the time when that part of packet
$p$ reaches $\dest{p}$). (Equivalently, this can be seen as a continuous
process, in which packet~$p$, at every time step $\tau\geq \release{p}$, incurs
a latency of $(1-X_{\tau}) \cdot \weight{p}$, where $X_{\tau}$ is the fraction
of $p$ that had been delivered before time $\tau$.) 

In particular, when the algorithm transmits packet~$p$ through the link that
connects $\source{p}$ and $\dest{p}$, the weighted latency of $p$ is $\weight{p}
\cdot \fixedDist{p}$. On the other hand, routing packet~$p$ via path $\source{p}
- t - r - \dest{p}$, where $(t,r)\in \Edges{p}$ incurs weighted latency of
$\weight{p} \cdot (\dist{\source{p}, t} + \dist{t,r} + \dist{r,\dest{p}}$.  

The goal of the algorithm is to deliver all packets and minimize the total
weighted (fractional) latency of its schedule. The overall performance of the
algorithm is measured with the standard notion of the competitive ratio, defined
as the worst-case \ALG-to-\OPT cost ratio, where \OPT is the optimal offline
solution with limited transmission speed.

\section{Online Scheduling Algorithm}
\label{sec:algo}

In this section we present our online scheduling algorithm \ALG, and will defer
its competitive analysis to the next section. Algorithm \ALG comprises two
natural components. A scheduler for packets in the reconfigurable network, which
relies on the repeated computation of \emph{stable matchings}, and a dispatcher
which upon arrival of a packet decides whether the packets will use a direct
connection or the reconfigurable links. In the latter case, the dispatcher
\emph{assigns} packet~$p$ to some edge (i.e., a transmitter-receiver pair) that
connects source and destination of $p$.

More precisely, the dispatcher attempts to minimize the weighted latency
increase caused by~$p$ (the latency of~$p$, and the latencies of other packets).
To this end, it needs to account for the different transmission times in the
reconfigurable part of the network. The idea is to split packets into
\emph{chucks}, which can be transmitted in a single step (the size of a chunk
depends on the delay of the assigned edge). The dispatcher is formally defined
in \lref{Section}{sec:dispatcher}. The scheduler then greedily chooses the
subset of chunks to be transmitted at each step. The set of edges associated
with each chunk forms a stable matching. This process is presented in details in
\lref{Section}{sec:matching}.

\subsection{Stable Matching and Blocking}

A matching $M$ is stable with respect to symmetric weights $w$ if for any edge~$e\not\in
M$, there exists edge~$e' \in M$ adjacent to $e$ such that $\weight{e'} \geq
\weight{e}$. We say that edge~$e'$ blocks edge~$e$. The scheduler at each time
$\tau$, transmits a set of packets whose assigned edges form a stable matching.
We will say that a chunk~$c$ blocks another chunk~$c'$ when $c$ is transmitted
at time $\tau$ and $\weight{c} \geq \weight{c'}$, and edges assigned to $c$ and
$c'$ share a transmitter or a receiver.

\subsection{Dispatcher}
\label{sec:dispatcher}
At time $\tau$, the dispatcher handles packets that arrived since time $\tau-1$.
The packets are processed one by one. We will say that a packet $p'$ arrived
before $p$ if $\release{p'} \leq \tau - 1$ or $\release{p'} = \tau$ and $p'$ was
already handled by the dispatcher. Each packet assigned to the reconfigurable is
split into chunks. For chuck $c$ we denote by $p(c)$ the packet whose part is
chunk~$c$. We say that a chunk is \emph{pending} if it has not been transmitted
through the reconfigurable network. For a set of chunks $C$ we denote the total
weight of chunks in $C$ by $\Weight{C}$. 

For each packet~$p$, let~$\before{p}$ be the set of chunks of packets that
arrived before $p$ in the input sequence and are pending at time when $p$ is
processed. We define the \emph{impact} of $p$ as the weighted latency of chunks
from $\before{p}$ that are blocked by (chunks of) $p$ plus the weighted latency
of (chunks of) $p$ incurred in rounds when $p$ were blocked by a chunk from
$\before{p}$ or its own chunk. In particular, if packet~$p$ is transmitted
through the fixed network, its impact is just the weighted latency of $p$, that
is $\weight{p} \cdot \fixedDist{p}$.

\begin{figure*}[t]
    \centering
    \begin{minipage}[c]{0.4\linewidth}
        \tikzstyle{circle}=[fill=gray, draw=black, shape=circle, scale=0.5]
\begin{tikzpicture}
    \node [circle, label={left:$s_1$}]  (0) at (-2.5, 3.5) {};
    \node [circle, label={left:$s_2$}]  (1) at (-2.5, 2.25) {};
    \node [circle, label={right:$d_1$}] (2) at (-0.5, 4) {};
    \node [circle, label={right:$d_2$}] (3) at (-0.5, 2.75) {};
    \node [circle, label={right:$d_3$}] (4) at (-0.5, 1.5) {};
    \node [circle, label={left:$s_1$}]  (5) at (1, 3.5) {};
    \node [circle, label={left:$s_2$}]  (6) at (1, 2.25) {};
    \node [circle, label={right:$d_1$}] (7) at (3, 4) {};
    \node [circle, label={right:$d_2$}] (8) at (3, 2.75) {};
    \node [circle, label={right:$d_3$}] (9) at (3, 1.5) {};
    \node (10) at (-1.5, 1) {packets $\Pi$};
    \node (11) at (2, 1) {packets $\Pi'$};

    \draw          (0) to node[anchor=south] {$p_1$} (2);
    \draw [dashed] (0) to node[anchor=south] {$p_2$} (3);
    \draw          (1) to node[anchor=south] {$p_3$} (3);
    \draw [dashed] (1) to                            (4);
    \draw [dashed] (5) to node[anchor=south] {$p_1$} (7);
    \draw          (5) to node[anchor=south] {$p_2$} (8);
    \draw [dashed] (6) to node[anchor=south] {$p_3$} (8);
    \draw          (6) to node[anchor=south] {$p_4$} (9);
\end{tikzpicture}
    \end{minipage}
    \quad\quad\quad\quad
    \begin{minipage}[c]{0.4\linewidth}
    \begin{center}
    \begin{tabular}{ c | c | c | r }
        packet & path & weight &  \multicolumn{1}{c}{impact} \\ 
        \hline
        $p_1$ & $s_1 \to d_1$ & $1$ & $\weight{p_1} = 1$ \\
        $p_2$ & $s_1 \to d_2$ & $2$ & $\weight{p_2} = 2$ \\
        $p_3$ & $s_2 \to d_2$ & $3$ & $\weight{p_2} + \weight{p_3} = 5$ \\
        \multicolumn{4}{c}{} \\ 
        packet & path & weight & \multicolumn{1}{c}{impact} \\ 
        \hline
        $p_1$ & $s_1 \to d_1$ & $1$ & $\weight{p_1}  = 1$ \\
        $p_2$ & $s_1 \to d_2$ & $2$ & $\weight{p_1} + \weight{p_2} = 3$ \\
        $p_3$ & $s_2 \to d_2$ & $3$ & $\weight{p_3}  = 3$ \\
        $p_4$ & $s_2 \to d_3$ & $4$ & $\weight{p_3} + \weight{p_4} = 7$ \\  
    \end{tabular}
    \end{center}
    \end{minipage}
    \caption{The figure shows a graph and two inputs (sets of packets). In the
    graph, for each source, there is exactly one transmitter attached to it and
    for each destination, there is exactly one receiver (the transmitters and
    receivers are omitted on the picture). The label above each edge is the
    (only) packet that might use this edge. Solid edges mark the stable matching
    (assuming the weight of an edge is the weight of the packet it can transmit)
    that would be transmitted if no more packets arrived. Upon arrival of a new
    packet $p_4$, the stable matching changes. As a result, packet $p_2$ is not
    blocked by $p_3$ and $p_2$ blocks $p_1$.}
    \label{fig:block} 
\end{figure*}

Ideally, when packet~$p$ arrives, we would like to minimize the impact of this
packet. This is, however, impossible to compute online: when the stable matching
changes as more packets arrive, the impact of packet~$p$ might change as well,
although the set $\before{p}$ does not change (it is a property of the input
sequence, not the algorithm). An example of such situation is depicted on
\lref{Figure}{fig:block}.

Instead, the algorithm minimizes the worst-case impact of $p$. Namely, for each
edge~$e\in\Edges{p}$, it computes, assuming that $p$ is assigned to $e$, how
many chunks from $\before{p}$ might block $p$ in the future and how many chunks
from $\before{p}$ might be blocked by $p$. Then, the algorithm minimizes the
worst-cast impact by assigning $p$ to either the edge from the reconfigurable
network to the direct fixed link between source and destination of $p$ (only if
such link exists).

Formally, when packet~$p$ is assigned to edge~$e$, it is split into $\dist{e}$
chunks, each of size $1/\dist{e}$ and weight $\weight{p} / \dist{e}$. Let
$\chunks{p}{e}$ be the set of these chunks. Let $\adj{p}{e}$ be the set of
chunks from $\before{p}$ that are assigned to use edge adjacent to~$e$. We
partition the set $\adj{p}{e}$ into two disjoint subsets: $\heavier{p}{e}$
containing those chunks that may delay $\chunks{p}{e}$ (i.e., at least as heavy
as $\weight{p}/ \dist{e}$) and $\lighter{p}{e}$ of chunks that might be delayed
by $\chunks{p}{e}$ (i.e., lighter than $\weight{p} / \dist{e}$). Note that these
definitions require that from two chunks of the same weight, the chunk of the
earlier arriving packet is preferred. This will be preserved by the scheduler in
the next section.

The worst-case impact of $p$ assigned to $e$ is then 
$
    \impE{p}{e} 
    = \weight{p} \cdot \left ( \dist{u} + \frac{\dist{e}+1}{2} + \dist{v} \right ) 
    + \weight{p} \cdot | \heavier{p}{e} |
    + \dist{e} \cdot \Weight{\lighter{p}{e}} \xspace.
$
The first summand is the weighted latency of chunks of $p$ (note that chunks
$\chunks{p}{e}$ delay each other). The remaining two summands account for the
latency increase coming from $p$ interacting with other chunks: i.e., they count
the number of chunks from $\adj{p}{e}$ that might block $p$, and the weighted
latency of packets from $\adj{p}{e}$ that may be blocked by $p$ (all
$|\chunks{p}{e}| = \dist{e}$ chunks of $p$ might block $\lighter{p}{e}$). 

Let $e = \arg\min_{e'\in \Edges{p}} \impE{p}{e'}$ be the edge that minimizes the
worst-case impact of $p$ among all edges of the reconfigurable network. If there
exists a link $e_\ell = (\source{p}, \dest{p})$ in $E_{\ell}$, and if the
weighted latency of sending $p$ through $e_\ell$ is smaller than the worst-case
impact of $p$ assigned to $e$ (i.e, $\weight{p}\cdot \fixedDist{p} \leq
\impE{p}{e}$), packet~$p$ is assigned to edge~$e_\ell$; otherwise, packet~$p$ is
assigned to edge~$e$.

If packet~$p$ is not transmitted via a direct source-destination link, the edge
$\edge{p}$, which will eventually transmit packet~$p$ is fixed. In the remaining
part of the paper we will use $\Heavier{p}$ and $\Lighter{p}$ to denote the
corresponding terms for the edge $\edge{p}$, that is, $\heavier{p}{\edge{p}}$
and $\lighter{p}{\edge{p}}$, respectively.

\subsection{Scheduler}
\label{sec:matching}

We describe how at time $\tau$ packets released until time $\tau$ are
transmitted through the reconfigurable network. To this end, we construct the
set $\match{\tau}$ of chunks that will be transmitted in the interval $[\tau,
\tau+1)$. The set of edges used by chunks from $\match{\tau}$ forms a stable
matching. We assume that each packet~$p$ is already assigned to an edge
$\edge{p}$ and split into chunks such that a chunk~$c$ assigned to edge~$e$ can
be transmitted in a single step, i.e., $\size{c} = 1 / \dist{e}$. For a chunk
$c$, by $p(c)$ we denote the packet whose part is $c$. The weight of $c$ is then
$\weight{c} = \weight{p} \cdot \size{c}$. 

The stable matching $\match{\tau}$ is constructed greedily. Initially, the set
$\match{\tau}$ is empty. Then, for each pending chunk~$c$, in order of
decreasing weights and increasing arrival times, if both endpoints of~$\edge{c}$
are free (i.e., chunks from $\match{\tau}$ do not use edges adjacent to
$\edge{c}$), chunk~$c$ becomes an element of $\match{\tau}$ (it will be
transmitted). Otherwise, if at least one of endpoints of $\edge{c}$ is already
busy, then $c$ is not transmitted. Observe that, due to our ordering with
decreasing weights, $c'$ blocks $c$ as $\weight{c'} \geq \weight{c}$. When the
algorithm processes all chunks, the matching $\match{\tau}$ is transmitted.

\section{Competitive Analysis}
\label{sec:analysis}

In this section we prove that our algorithm $\ALG$ is
$2\cdot(2+\varepsilon)/\varepsilon$-competitive given a $(2+\varepsilon)$
speedup. In the analysis, instead of empowering the algorithm, we limit the
capabilities of the optimum algorithm: For $\varepsilon\geq 0$ its transmission
can take time at most $1/(2+\varepsilon)$ in a single step.
Although the schedule of the algorithm is non-migratory (a packet is assigned to
and transmitted via exactly one path in $G$), the result holds against an
optimal solution that is preemptive and migratory.

\subsection{Linear Program Relaxation}

\begin{figure*}
    \begin{align*}
        \text{min. \quad}  & \sum_{p\in\Packets} \sum_{e\in \Edges{p}} 
        \sum_{\tau\geq \release{p}}  \weight{p} \cdot x_{p,e,\tau} 
            \cdot \left ( \tau + \Dist{e} - \release{p} \right ) 
            + \sum_{p\in \PacketsLinks} \weight{p} \cdot y_p \cdot \fixedDist{p} \\
        \text{s.t. \quad} 
        & \sum_{e\in \Edges{p}} \sum_{\tau \geq \release{p}} x_{p,e,\tau} 
            + y_p \geq 1 &  \text{for all } p\in \PacketsLinks \\
        & \sum_{e\in \Edges{p}} \sum_{\tau \geq \release{p}} x_{p,e,\tau} 
            \geq 1 &  \text{for all } p\in \Packets \setminus \PacketsLinks \\
        & \sum_{r\in R(t)} \sum_{p\in P(\tau) :{(t,r)\in \Edges{p}}} 
        \dist{t,r} \cdot x_{p,(t,r),\tau} \leq \frac{1}{2+\varepsilon} & \text{for
        all } \tau,\ t\in \trans \\
        & \sum_{t\in T(r)} \sum_{p\in P(\tau) :{(t,r)\in \Edges{p}}} 
        \dist{t,r} \cdot x_{p,(t,r),\tau} \leq \frac{1}{2+\varepsilon} & \text{for
        all } \tau,\ r \in \receiver 
    \end{align*}
    \caption{Linear program \primal describing all feasible solutions with
    reduced transmission speed in the reconfigurable network. We omit nonnegativity
    constrains of all variables.}\label{fig:primal} 
    \begin{align*}
        \text{max. \quad}  
        & \sum_{p\in\Packets} \alpha_{p} - \frac{1}{2+\varepsilon} \left(\sum_{t\in \trans } 
        \sum_{\tau} \beta_{t,\tau}  + \sum_{r\in \receiver } \sum_{\tau} \beta_{r,\tau} \right) \\
        \text{s.t. \quad} 
        & \alpha_{p} - \dist{e} \cdot (\beta_{t,\tau} + \beta_{r,\tau}) 
        \leq \weight{p} \cdot (\tau + \Dist{e} - \release{p})  
        & \text{ for all } p\in\Packets, e=(t,r)\in \Edges{p}, \tau\geq \release{p} \\
        & \alpha_{p} \leq \weight{p} \cdot \fixedDist{p}  & \text{ for all } p \in \PacketsLinks
    \end{align*}
    \caption{Linear program \dual dual to \primal. We omit nonnegativity constrains of all variables.}
    \label{fig:dual}
\end{figure*}

For our analysis we will rely on the formulation of primal and dual linear
programs (containing all feasible solutions transmitting packets with speed
$1/(2+\varepsilon)$), to which we will be able to charge the costs of our online
algorithm. This will eventually allow us to upper bound the competitive ratio
(how far our algorithm is off from a best possible offline solution).

For packet $p\in \Packets$, edge~$e=(t,r)\in\Edges{p}$ and time $\tau\geq
\release{p}$ we introduce a variable $x_{p,e,\tau}$ interpreted as a fraction of
packet~$p$ that is sent through the path $\source{p}-t-r-\dest{p}$ at time
$\tau$. Note that, this transmission takes time $\Dist{e} = \dist{\source{t}, t}
+ \dist{e} + \dist{r,\dest{r}}$, which incurs weighted latency of $\weight{p}
\cdot x_{p,e,\tau}\cdot (\tau + \Dist{e} - \release{p})$. To model fixed direct
links between sources and destinations, for each packet~$p$ we introduce
variable $y_{p}$, which is interpreted as the amount sent through this direct
connection. The weighted latency of this transmission is then $\weight{p}\cdot
\fixedDist{p} \cdot y_p$.

For time $\tau$, let $P(\tau)$ be the set of packets released earlier than
$\tau$. The linear program \primal shown in \lref{Figure}{fig:primal} describes
the set of feasible solutions (in particular, the optimal solution in a reaource
augmentation model).

Note that the number of variables and constraints is potentially infinite, but
it is sufficient to consider only $\tau$ smaller than $\max_{p\in\Packets}
\release{p} + |\Packets| \cdot \max_{e\in E} \Dist{e}$. This is because if there
is any pending packet, then any (reasonable) algorithm transmits at least one of
them and transmitting packets one by one takes time at most $|\Packets| \cdot
\max_{e\in E} \Dist{e}$.

The first and the second sets of constraints force transmitting all packets
either through the reconfigurable or the fixed network (if the latter is
possible). The remaining two sets of constraints ensure that at any time, sets
of edges corresponding to transmitted packets form a matching in the
reconfigurable network and limit the transmission time to $1/(2+\varepsilon)$
(recall that sending amount $s$ through edge~$e$ takes time $s\cdot \dist{e}$).

However, we note that our algorithm will not give feasible solutions for the
primal LP, as we are relying on resource augmentation. Hence, we next formulate
the corresponding dual. The program \dual dual to \primal is shown in
\lref{Figure}{fig:dual}.

\subsection{Solution to Dual Program}
\label{sec:dualSolution}

From the weak duality, the value of a feasible solution to program \dual is a
lower bound on the cost of the optimal solution to~\primal. We utilize this to
prove that algorithm $\ALG$ is competitive. For the sake of analysis, we
construct a feasible dual solution whose cost can be related to the cost of
$\ALG$.

In the solution to \dual used throughout the analysis, for packet $p\in
\Packets$ we set the value of $\alpha_{p}$ to the worst-case impact of~$p$
estimated at the arrival of this packet: If packet~$p$ was transmitted through
the direct source-destination link, we set $\alpha_{p} = \fixedDist{p}$.
Otherwise, if packet~$p$ was sent through edge~$\edge{p}$ in the reconfigurable
network, the value of $\alpha_p$ is set to $\alpha_{p} = \impE{p}{\edge{p}}$.

For time $\tau$, transmitter $t\in T$ and receiver $r\in R$, let $C_{t,\tau}$
and $C_{r,\tau}$ be the set of all chunks assigned to use the edge adjacent to
$t$ and $r$, respectively, that have not reached their destination until
time~$\tau$. We set $\beta_{t,\tau} = \Weight{C_{t,\tau}}$ and
$\beta_{r,\tau}=\Weight{C_{r,\tau}}$ to the total weight of chunks from
corresponding sets $C$.

We will abuse the notation and use \dual for both dual linear program and its
solution defined in this section.

\subsection{\ALG-to-Dual Ratio}

The goal of this section is to relate the cost of \ALG to the objective value of
the dual solution. First, we show that the latency accumulated in $\beta$
variables is at most twice the cost of the algorithm. Second, we define a cost
charging scheme from the weighted latency of \ALG to variables $\alpha$. Third,
by jointly considering these two relations, we get that the cost of the
algorithm is at most $(2+\varepsilon)/ \varepsilon$ times the value of the dual
solution.

\begin{lemma}
\label{lem:beta_charge}
    $\ALG \geq \sum_{t\in T} \sum_{\tau} \beta_{t,\tau} = \sum_{r\in R}
    \sum_{\tau} \beta_{r,\tau}$.
\end{lemma}

\begin{proof}
The packets that use direct connections incur (positive) latency, but are not
counted by the $\beta$ variables. Therefore, it is sufficient to prove that the
sum of transmitters' $\beta$ variables as well as the sum of receivers' $\beta$
variables equals the weighted latency of packets (chunks) that are transmitted
via the reconfigurable network.

Fix a chunk~$c$ of packet~$p$ that was transmitted via reconfigurable network.
Let $\act{c}$ denote the period when $c$ was active, that is, all the times
$\tau$ from $\release{p}$ until the time $c$ reaches $\dest{p}$. For any time
$\tau$ in $\act{c}$, chunk~$c$ incurs cost $\weight{c}$. Recall that chunk~$c$
is assigned to exactly one edge $\edge{p(c)}$ from~$\Edges{p(c)}$ and thus to
exactly one transmitter $t$ and receiver $r$ (the endpoints of edge~$e$). Hence,
for each $\tau\in\act{c}$, it holds that $c \in C_{t,\tau}$ and $c\in
C_{r,\tau}$, so for each time $\tau\in \act{c}$, $\weight{c}$ is counted towards
$\beta_{t,\tau}$ and $\beta_{r,\tau}$. These are the only $\beta$ variables that
count the latency of $c$ at transmission step $\tau$. The lemma follows by
summing over all packets and their chunks in the input.
\end{proof}

\noindent
\textbf{\ALG-to-$\alpha$'s charging scheme.}
In this section we charge the cost of the algorithm (the weighted latencies) to
packets. The goal is to show, that each packet is charged at most the value of
the corresponding $\alpha$ variable. This will let us relate the cost of the
algorithm to the dual solution.

Fix packet~$p$. If $p$ was transmitted via the fixed network, then we simply
charge its total latency to $p$ itself. Otherwise, $p$ was split into several
chunks. The chunks of $p$ might delay each other as edge $\edge{p}$ can transmit
just one of them in a single transmission. Therefore, we will focus on a single
chunk of packet~$p$ and charge its latency to other packets or packet~$p$. 

Let $c\in\chunks{p}{\edge{p}}$ be a chunk of $p$. It incurs weighted latency of
$\weight{c}$ for each time $\tau \in \act{c}$ before it reached $\dest{p}$. When
$c$ is being transmitted through any edge of the graph, we simply charge
$\weight{c}$ to $p$. It remains to charge latency of $c$ when it was waiting in
the transmitter's queue. For each such time~$\tau$, there exists another chunk
that blocked $c$. Let $B$ be the set of all chunks that blocked $c$.

For each chunk~$c' \in B$, if $c'$ and $c$ are parts of the same packet~$p$, the
latency $\weight{c}$ is charged, again, to $p$. Otherwise, we charge
$\weight{c}$ to $p$ or $p'=p(c)$ depending on which of these two packets arrived
later. If $\release{p} < \release{p'}$, the latency of $\weight{c}$ is charged
to $p'$. Note that $c'$ is heavier than $c$ and hence $c \in \Lighter{p'}$. If
$\release{p'} < \release{p}$ we charge $\weight{c}$ to packet~$p$. If this is
the case, it holds that $c' \in \Heavier{p}$.

In the following lemma we bound the charges received by each packet.

\begin{lemma} 
\label{lem:alpha_charge} 
    For packet~$p$, let $c_{p}$ be the weighted latency charged to $p$. Then, it
    holds that $c_p \leq \alpha_{p}$.
\end{lemma}

\begin{proof}
Fix packet~$p$. If $p$ is sent via the fixed network, the only charge it
receives is $\weight{p}\cdot \fixedDist{p} = \alpha_p$. Otherwise, the charges
received by $p$ are threefold:

\begin{itemize}

\item First, for each of its chunks, packet~$p$ receives a charge of
$\weight{c}$ for every time $\tau$ when $c$ was not blocked (i.e., transmitted
via any edge) or blocked by other chunk of $p$. The $i$-th (for $i\in \{1, 2,
\ldots, \dist{\edge{p}} \}$) delivered chunk of $p$ charges $\weight{c}\cdot
\left(\dist{\source{p},t} + i + \dist{r,\dest{p}} \right )$. In total the
latency charged in this case is equal to 
\begin{align*}
    & \sum_{i = 1}^{\dist{\edge{p}}} \frac{\weight{p}}{\dist{\edge{p}}}\cdot
    \left(\dist{\source{p},t} + i + \dist{r,\dest{p}} \right ) \\
    & =  \weight{p}\cdot
        \left(\dist{\source{p},t} + \frac{\dist{\edge{p}} +1}{2} + \dist{r,\dest{p}} \right ) .
\end{align*}

\item Second, when $c$ blocked chunk~$c'$ of a packet that arrived earlier than
$p$, it receives charge~$\weight{c'}$. This charge can be received only from
packets in set $\Lighter{p}$. Note that $c'$ is delayed by all $\dist{\edge{p}}$
chunks of $p$.

\item Third, when $c$ is blocked by some other chunk~$c'$ of packet that arrived
earlier than $p$, packet~$p$ receives a charge of $\weight{c}$. In this case,
$c' \in \Heavier{p}$.

\end{itemize}

Combining all three cases, we obtain that the latency charged to $p$ is at most
\begin{align*}
    c_p \leq \weight{p}\cdot
    \left(\dist{\source{p},t} + \frac{\dist{\edge{p}} +1}{2} + \dist{r,\dest{p}} \right ) + 
    \\  \weight{p} \cdot |\Heavier{p}| + \dist{e} \cdot \Weight{\Lighter{p}} = \alpha_p
\end{align*}
\end{proof}

\begin{lemma}
    \label{lem:algDualRatio}
    For any $\varepsilon > 0$ it holds that $\ALG \leq (2+\varepsilon)/
    \varepsilon \cdot \dual$.
\end{lemma}
\begin{proof}
Fix $\varepsilon > 0$. If we sum the guarantees from
\lref{Lemma}{lem:alpha_charge} over all packets $p \in \Packets$, we obtain
$\ALG \leq \sum_{p} \alpha_{p}$. This combined with
\lref{Lemma}{lem:beta_charge} immediately yields the lemma, as
\begin{align*} 
    \dual &
    \geq \sum_{p\in\Packets} \alpha_{p} - \frac{1}{2+\varepsilon} \cdot\left(
    \sum_{t \in  \trans} \sum_{\tau} \beta_{t,\tau} +
    \sum_{r \in \receiver} \sum_{\tau} \beta_{r,\tau} \right)  \\
    & \geq \ALG - \frac{2}{2+\varepsilon}\cdot \ALG 
    = \frac{\varepsilon}{2+\varepsilon}\cdot \ALG.
\end{align*}
\end{proof}

\subsection{Dual-to-\OPT Ratio}

By weak duality, the value of any feasible solution to \dual is a lower bound on
the cost of \OPT. Our assignment of $\alpha$ and $\beta$ variables does not
necessarily constitute a feasible solution to \dual, that is, some constraints
might be violated. However, these constraints are "almost feasible", i.e., the
solution obtained by halving each variable is feasible. Therefore, the value of
our dual solution is at most twice the optimum, which together with the results
of the previous section, will lead to the bound on the competitive ratio of
\ALG.

The following lemma shows that constraints in \dual corresponding to packet~$p$
and edge~$e$ are violated by a factor of $2$ if instead of $\alpha_p$ we take
the value of precomputed impact of $p$ assigned to edge~$e$.
\begin{lemma}
    \label{lem:impBound}
    For packet~$p$, edge~$e=(t,r)\in \Edges{p}$ and time $\tau\geq \release{p}$
    it holds that
    \[
        \impE{p}{e} - \dist{e} \cdot \left( \beta_{t,\tau} + \beta_{r,\tau} \right)
        \leq 2\cdot \weight{p}\cdot\left(t + \Dist{e} - \release{p}\right).
    \]
\end{lemma}
\begin{proof}
Fix packet~$p$, edge~$e=(t,r)\in \Edges{p}$ and time $\tau\geq \release{p}$. We
start with proving that 
\begin{align}
    \label{eq:deltaEdge}
    L & := \weight{p}\cdot |\heavier{p}{e} | + \dist{e} \cdot \Weight{\lighter{p}{e}} \\
    & \leq \dist{e} \cdot (\beta_{t,\tau} + \beta_{r,\tau}) 
    + 2 \cdot \weight{p} \cdot (\tau - \release{p}). \nonumber
\end{align} 
To this end observe that the contribution of a single chunk~$c\in\adj{p}{e}$
(i.e., chunk that uses edge adjacent to $e$) towards $L$ is at most
\begin{equation} 
    \label{eq:chunk_to_L}
    \min\left ( \dist{e} \cdot \weight{c}, \weight{p} \right) .
\end{equation}
Two nontrivial relations follow directly from definitions of sets
$\heavier{p}{e}$ and $\lighter{p}{e}$. If $c\in\heavier{p}{e}$, then $\weight{p}
\leq \weight{c} \cdot \dist{e}$ and when $c\in\lighter{p}{e}$, then $\dist{e}
\cdot \weight{c} \leq \weight{p}$.

Let $P$ be the set of those chunks in $\adj{p}{e}$ that have not reached their
destination by time $\tau$. By~\eqref{eq:chunk_to_L}, the contribution of chunks
from $P$ towards $L$ is at most $\dist{e} \cdot \Weight{P} \leq \dist{e}\cdot
\left( \beta_{t,\tau} + \beta_{r,\tau} \right)$.

It remains to bound the contribution of packets in $Q=\adj{p}{e} \setminus P$.
Again, by \eqref{eq:chunk_to_L}, chunk~$c\in Q$ contributes at most $\weight{p}$
towards $L$. The proof of \eqref{eq:deltaEdge} is concluded by observing that
the set $Q$ contains at most $2\cdot (\tau-\release{p})$ chunks as endpoints of
edge~$e$ transmit at most one chunk in each transmission step. 

The lemma follows by combining inequality \eqref{eq:deltaEdge}, the fact that
$\dist{e} \geq 1$, and the definition of $\impE{p}{e}$ :
\begin{align*}
    \impE{p}{e}  
    & = \weight{p} \cdot \left( \dist{t} + \frac{\dist{e}+1}{2} + \dist{r} \right) \\
    & + \weight{p} \cdot | \heavier{p}{e} |
    + \dist{e} \cdot \Weight{\lighter{p}{e}} \\
    & \leq  \weight{p} \cdot \Dist{e}
    + \dist{e} \cdot (\beta_{t,\tau} + \beta_{r,\tau}) 
    + 2 \cdot \weight{p} \cdot (\tau - \release{p})\\
    & < 2\cdot \weight{p} \cdot \left (\tau + \Dist{e} - \release{p} \right ) 
    + \dist{e} \cdot (\beta_{t,\tau} + \beta_{r,\tau}).
\end{align*}
\end{proof}

In the next lemma we combine weak duality to relate the value of our dual
solution to the value of optimum.
\begin{lemma}
    \label{lem:dualFeasible}
    The value of our solution to dual program is at most twice the value of
    optimal solution, i.e., $\dual\leq 2 \cdot \OPT$.
\end{lemma}
\begin{proof}
We prove that the solution to \dual obtained by halving each variable $\alpha$
and $\beta$ is feasible. To this end we show that the solution defined in
\lref{Section}{sec:dualSolution} violates constraints by a factor at most $2$.

First, for packet $p\in\PacketsLinks$, the constraint corresponding to primal
variable $y_p$ (i.e., routing through the fixed network) is not violated as
$\alpha_p$ minimizes the impact of routing $p$ through any path, in particular,
the direct source-destination link and hence $\alpha_p \leq \weight{p} \cdot
\fixedDist{p}$.

Second, fix the dual constraint corresponding to primal variable $x_{p,e,\tau}$
for packet $p\in\Packets$, edge~$e = (t,r)\in\Edges{p}$ and time $\tau\geq
\release{p}$. 

Applying the definition of $\alpha_p$ and \lref{Lemma}{lem:impBound} to the
left-hand side of the dual constraint we obtain:
\begin{align*}
    \alpha_p - \dist{e}\cdot \left( \beta_{t,\tau} + \beta_{r,\tau} \right) 
    & \leq \impE{p}{e} - \dist{e}\cdot \left( \beta_{t,\tau} + \beta_{r,\tau} \right) \\
    & \leq 2\cdot \weight{p}\cdot\left(\tau+\Dist{e}-\release{p}\right).
\end{align*}

Therefore, the dual solution created by halving each variable is feasible. The
lemma follows from weak duality.
\end{proof}

By combining \lref{Lemma}{lem:algDualRatio} and \lref{Lemma}{lem:dualFeasible},
we derived the following theorem:
\begin{theorem}
    For any $\varepsilon > 0$, $\ALG$ is $2\cdot \left({2}/{\varepsilon} + 1
\right)$-competitive with speedup $(2+\varepsilon)$. That is, for any
input~$\Inp$, if \ALG works $2+\varepsilon$ times faster than the optimal
offline algorithm \OPT, the cost of \ALG is bounded by $\ALG \leq 2\cdot
\left({2}/{\varepsilon} + 1 \right) \cdot \OPT$.
\end{theorem}

\section{Related Work}
\label{sec:relwork}

Reconfigurable optical topologies~\cite{zhou2012mirror, kandula2009flyways,
rotornet, opera, firefly, megaswitch, quartz,
chen2014osa,projector,ballani2020sirius} have recently received much attention
in the literature as an alternative to traditional static datacenter topologies
~\cite{clos,jupiter,f10,bcube,mdcube,xpander,jellyfish}. It has been
demonstrated that already demand-oblivious reconfigurable topologies can deliver
unprecedented bandwidth efficiency~\cite{rotornet,opera,ballani2020sirius}. By
additionally exploiting the typical skewed and bursty structure of traffic
workloads~\cite{tracecomplexity,benson2010network,facebook,kandula2009nature,mogul2012we,DBLP:journals/cn/ZouW0HCLXH14,
datacenter_burstiness}, demand-aware reconfigurable topologies can be further
optimized, e.g., toward elephant flows. To this end, existing demand-aware
networks are based on traffic matrix predictions~\cite{proteus,osa,dan} or even
support per-flow or ``per-packet''
reconfigurations~\cite{projector,megaswitch,venkatakrishnan2018costly,helios,mordia,cthrough,splaynet}.
Our focus on this paper is on the latter, and we consider fine-grained
scheduling algorithms.

While most existing systems are mainly evaluated empirically, some also come
with formal performance guarantees. However, to the best of our knowledge,
besides some notable exceptions which however focus on single-tier
models~\cite{perf20bmatch,avin2019renets,eclipse,dinitz2020scheduling,schwartz2019online},
hardly anything is known about the achievable competitive ratio by online packet
scheduling algorithms. In particular, our paper is motivated by the multi-tier
ProjecToR architecture~\cite{projector}, to which our analysis also applies.

Our model and result generalizes existing work on competitive switch scheduling
\cite{mckeown1999islip,chuang1999matching}: In classic switch scheduling,
packets arriving at a switch need to be moved from the input buffer to the
output buffer, and in each time step, the input buffers and all their output
buffers must form a bipartite matching. A striking result of Chuang, Goel,
McKeown, and Prabhakar showed that a switch using input/output queueing with a
speed-up of 2 can simulate a switch that uses pure output queueing
\cite{chuang1999matching}. Our model generalizes this problem to a multi-tier
problem, and we use a novel primal-dual charging scheme.

Venkatakrishnan et al.~\cite{eclipse} initiated the study of an offline
scheduling variant of the circuit switch scheduling problem, motivated by
reconfigurable datacenters. They consider a setting in which demand matrix
entries are small, and analyze a greedy algorithm achieving an (almost) tight
approximation guarantee. In particular, their model allows to account for
reconfiguration delays, which are not captured by traditional crossbar switch
scheduling algorithms, e.g., relying on centralized Birkhoff-von- Neumann
decomposition schedulers~\cite{mckeown1999achieving}. Schwartz et al.
\cite{schwartz2019online} recently presented online greedy algorithms for this
problem, achieving a provable competitive ratio over time. This line of research
however is technically fairly different from ours: the authors consider a
maximization problem, aiming to maximize the total data transmission for a
certain time window, whereas in our model, we aim to minimize completion times
(i.e., \emph{all} data needs to be transmitted).
Furthermore, while we consider a multi-tier network (inspired by architectures
such as~\cite{projector}), these works assume a complete bipartite graph. Last
but not least, we also support a simple form of hybrid architectures in our
model. Besides these differences in the model, our model differs significantly
from~\cite{schwartz2019online,eclipse} in terms of the used techniques. While
prior work relies, among other, on randomized rounding, we study an online
primal-dual approach.

In general, online primal-dual approaches have received much attention recently,
after the seminal work by Buchbinder and Naor~\cite{buchbinder2009design}.
Unlike much prior work in this area, we however do not use the online
primal-dual approach for the design of an algorithm, but only for its analysis.
In this regard, our approach is related to scheduling literature by Anand et al.
\cite{anand2012resource}, and the interesting work by Dinitz et al.
\cite{dinitz2020scheduling} on  reconfigurable networks. We generalize the
analysis of~\cite{anand2012resource} to a more general graph where we can have
conflicts at receives and transmitters. In~\cite{dinitz2020scheduling}, a model
is considered in which we are given an arbitrary graph; the demands are the
edges in the graph, and in each round, a vertex cover can be communicated: each
node can only send certain number of packets in one round. In contrast, we
consider bipartite graphs in which in each round, a \emph{matching} can be
transmitted, as it is supported (and hence motivated) by existing optical
technologies, both related to circuit switches
\cite{rotornet,flexspander,ballani2020sirius} and free-space optics
\cite{projector}. The optimization problems resulting from these models differ
significantly, requiring different algorithmic and analytic techniques.

\section{Conclusion}
\label{sec:conclusion}

We presented a competitive scheduling algorithm for reconfigurable datacenter
networks which generalizes classic switch scheduling problems and whose analysis
relies on a dual-fitting approach. We understand our work as a first step and
believe that it opens several interesting avenues for future research. In
particular, it would be interesting to study the optimality of bicriteria
scheduling algorithms. Furthermore, it would be interesting to explore
randomized scheduling algorithms.

{\balance
  \bibliographystyle{ieeetr} 
   \balance
\bibliography{literature}

\begin{thebibliography}{10}

\bibitem{singh2015jupiter}
A.~Singh, J.~Ong, A.~Agarwal, G.~Anderson, A.~Armistead, R.~Bannon, S.~Boving,
  G.~Desai, B.~Felderman, P.~Germano, {\em et~al.}, ``Jupiter rising: A decade
  of clos topologies and centralized control in google's datacenter network,''
  {\em ACM SIGCOMM computer communication review}, vol.~45, no.~4,
  pp.~183--197, 2015.

\bibitem{clos}
M.~Al-Fares, A.~Loukissas, and A.~Vahdat, ``A scalable, commodity data center
  network architecture,'' in {\em ACM SIGCOMM Computer Communication Review},
  vol.~38, pp.~63--74, ACM, 2008.

\bibitem{bcube}
C.~Guo, G.~Lu, D.~Li, H.~Wu, X.~Zhang, Y.~Shi, C.~Tian, Y.~Zhang, and S.~Lu,
  ``Bcube: a high performance, server-centric network architecture for modular
  data centers,'' {\em ACM SIGCOMM Computer Communication Review}, vol.~39,
  no.~4, pp.~63--74, 2009.

\bibitem{slimfly}
M.~Besta and T.~Hoefler, ``Slim fly: A cost effective low-diameter network
  topology,'' in {\em SC'14: Proceedings of the International Conference for
  High Performance Computing, Networking, Storage and Analysis}, pp.~348--359,
  IEEE, 2014.

\bibitem{singla2012jellyfish}
A.~Singla, C.-Y. Hong, L.~Popa, and P.~B. Godfrey, ``Jellyfish: Networking data
  centers randomly,'' in {\em Presented as part of the 9th $\{$USENIX$\}$
  Symposium on Networked Systems Design and Implementation ($\{$NSDI$\}$ 12)},
  pp.~225--238, 2012.

\bibitem{xpander}
S.~Kassing, A.~Valadarsky, G.~Shahaf, M.~Schapira, and A.~Singla, ``Beyond
  fat-trees without antennae, mirrors, and disco-balls,'' in {\em Proceedings
  of the Conference of the ACM Special Interest Group on Data Communication},
  pp.~281--294, ACM, 2017.

\bibitem{opera}
W.~M. Mellette, R.~Das, Y.~Guo, R.~McGuinness, A.~C. Snoeren, and G.~Porter,
  ``Expanding across time to deliver bandwidth efficiency and low latency,''
  {\em arXiv preprint arXiv:1903.12307}, 2019.

\bibitem{rotornet}
W.~M. Mellette, R.~McGuinness, A.~Roy, A.~Forencich, G.~Papen, A.~C. Snoeren,
  and G.~Porter, ``Rotornet: A scalable, low-complexity, optical datacenter
  network,'' in {\em Proceedings of the Conference of the ACM Special Interest
  Group on Data Communication}, pp.~267--280, ACM, 2017.

\bibitem{helios}
N.~Farrington, G.~Porter, S.~Radhakrishnan, H.~H. Bazzaz, V.~Subramanya,
  Y.~Fainman, G.~Papen, and A.~Vahdat, ``Helios: a hybrid electrical/optical
  switch architecture for modular data centers,'' {\em ACM SIGCOMM Computer
  Communication Review}, vol.~41, no.~4, pp.~339--350, 2011.

\bibitem{cthrough}
G.~Wang, D.~G. Andersen, M.~Kaminsky, K.~Papagiannaki, T.~Ng, M.~Kozuch, and
  M.~Ryan, ``c-through: Part-time optics in data centers,'' {\em ACM SIGCOMM
  Computer Communication Review}, vol.~41, no.~4, pp.~327--338, 2011.

\bibitem{projector}
M.~Ghobadi, R.~Mahajan, A.~Phanishayee, N.~Devanur, J.~Kulkarni, G.~Ranade,
  P.-A. Blanche, H.~Rastegarfar, M.~Glick, and D.~Kilper, ``Projector: Agile
  reconfigurable data center interconnect,'' in {\em Proceedings of the 2016
  ACM SIGCOMM Conference}, pp.~216--229, ACM, 2016.

\bibitem{avin2017demand}
C.~Avin, K.~Mondal, and S.~Schmid, ``Demand-aware network designs of bounded
  degree,'' {\em Distributed Computing}, pp.~1--15, 2017.

\bibitem{ballani2020sirius}
H.~Ballani, P.~Costa, R.~Behrendt, D.~Cletheroe, I.~Haller, K.~Jozwik,
  F.~Karinou, S.~Lange, K.~Shi, B.~Thomsen, {\em et~al.}, ``Sirius: A flat
  datacenter network with nanosecond optical switching,'' in {\em Proceedings
  of the Annual conference of the ACM Special Interest Group on Data
  Communication on the applications, technologies, architectures, and protocols
  for computer communication}, pp.~782--797, 2020.

\bibitem{splaynet}
S.~Schmid, C.~Avin, C.~Scheideler, M.~Borokhovich, B.~Haeupler, and Z.~Lotker,
  ``Splaynet: Towards locally self-adjusting networks,'' {\em IEEE/ACM
  Transactions on Networking (ToN)}, vol.~24, no.~3, pp.~1421--1433, 2016.

\bibitem{eclipse}
S.~Bojja~Venkatakrishnan, M.~Alizadeh, and P.~Viswanath, ``Costly circuits,
  submodular schedules and approximate carath{\'e}odory theorems,'' in {\em
  Proc. ACM SIGMETRICS}, pp.~75--88, 2016.

\bibitem{xweaver}
M.~Wang, Y.~Cui, S.~Xiao, X.~Wang, D.~Yang, K.~Chen, and J.~Zhu, ``Neural
  network meets dcn: Traffic-driven topology adaptation with deep learning,''
  {\em Proceedings of the ACM on Measurement and Analysis of Computing
  Systems}, vol.~2, no.~2, p.~26, 2018.

\bibitem{facebook}
A.~Roy, H.~Zeng, J.~Bagga, G.~Porter, and A.~C. Snoeren, ``Inside the social
  network's (datacenter) network,'' in {\em ACM SIGCOMM Computer Communication
  Review}, vol.~45, pp.~123--137, ACM, 2015.

\bibitem{benson2010network}
T.~Benson, A.~Akella, and D.~A. Maltz, ``Network traffic characteristics of
  data centers in the wild,'' in {\em Proceedings of the 10th ACM SIGCOMM
  conference on Internet measurement}, pp.~267--280, ACM, 2010.

\bibitem{tracecomplexity}
C.~Avin, M.~Ghobadi, C.~Griner, and S.~Schmid, ``On the complexity of traffic
  traces and implications,'' in {\em Proc. ACM SIGMETRICS}, 2020.

\bibitem{mckeown1999islip}
N.~McKeown, ``The islip scheduling algorithm for input-queued switches,'' {\em
  IEEE/ACM transactions on networking}, vol.~7, no.~2, pp.~188--201, 1999.

\bibitem{chuang1999matching}
S.-T. Chuang, A.~Goel, N.~McKeown, and B.~Prabhakar, ``Matching output queueing
  with a combined input/output-queued switch,'' {\em IEEE Journal on Selected
  Areas in Communications}, vol.~17, no.~6, pp.~1030--1039, 1999.

\bibitem{dinitz2020scheduling}
M.~{Dinitz} and B.~{Moseley}, ``Scheduling for weighted flow and completion
  times in reconfigurable networks,'' in {\em IEEE Conference on Computer
  Communications (INFOCOM)}, pp.~1043--1052, 2020.

\bibitem{gale1962college}
D.~Gale and L.~S. Shapley, ``College admissions and the stability of
  marriage,'' {\em The American Mathematical Monthly}, vol.~69, no.~1,
  pp.~9--15, 1962.

\bibitem{kalyanasundaram2000speed}
B.~Kalyanasundaram and K.~Pruhs, ``Speed is as powerful as clairvoyance,'' {\em
  Journal of the ACM (JACM)}, vol.~47, no.~4, pp.~617--643, 2000.

\bibitem{anand2012resource}
S.~Anand, N.~Garg, and A.~Kumar, ``Resource augmentation for weighted flow-time
  explained by dual fitting,'' in {\em Proceedings of the twenty-third annual
  ACM-SIAM symposium on Discrete Algorithms}, pp.~1228--1241, SIAM, 2012.

\bibitem{buchbinder2009design}
N.~Buchbinder and J.~Naor, {\em The design of competitive online algorithms via
  a primal-dual approach}.
\newblock Now Publishers Inc, 2009.

\bibitem{zhou2012mirror}
X.~Zhou, Z.~Zhang, Y.~Zhu, Y.~Li, S.~Kumar, A.~Vahdat, B.~Y. Zhao, and
  H.~Zheng, ``Mirror mirror on the ceiling: Flexible wireless links for data
  centers,'' {\em Proc. ACM SIGCOMM Computer Communication Review (CCR)},
  vol.~42, no.~4, pp.~443--454, 2012.

\bibitem{kandula2009flyways}
S.~Kandula, J.~Padhye, and P.~Bahl, ``Flyways to de-congest data center
  networks,'' in {\em Proc. ACM Workshop on Hot Topics in Networks (HotNets)},
  2009.

\bibitem{firefly}
N.~Hamedazimi, Z.~Qazi, H.~Gupta, V.~Sekar, S.~R. Das, J.~P. Longtin, H.~Shah,
  and A.~Tanwer, ``Firefly: A reconfigurable wireless data center fabric using
  free-space optics,'' in {\em ACM SIGCOMM Computer Communication Review},
  vol.~44, pp.~319--330, ACM, 2014.

\bibitem{megaswitch}
L.~Chen, K.~Chen, Z.~Zhu, M.~Yu, G.~Porter, C.~Qiao, and S.~Zhong, ``Enabling
  wide-spread communications on optical fabric with megaswitch,'' in {\em
  Proceedings of the 14th USENIX Conference on Networked Systems Design and
  Implementation}, NSDI'17, (USA), pp.~577--593, USENIX Association, 2017.

\bibitem{quartz}
Y.~J. Liu, P.~X. Gao, B.~Wong, and S.~Keshav, ``Quartz: A new design element
  for low-latency dcns,'' {\em SIGCOMM Comput. Commun. Rev.}, vol.~44,
  pp.~283--294, Aug. 2014.

\bibitem{chen2014osa}
K.~Chen, A.~Singla, A.~Singh, K.~Ramachandran, L.~Xu, Y.~Zhang, X.~Wen, and
  Y.~Chen, ``Osa: An optical switching architecture for data center networks
  with unprecedented flexibility,'' {\em IEEE/ACM Transactions on Networking
  (TON)}, vol.~22, no.~2, pp.~498--511, 2014.

\bibitem{jupiter}
A.~Singh, J.~Ong, A.~Agarwal, G.~Anderson, A.~Armistead, R.~Bannon, S.~Boving,
  G.~Desai, B.~Felderman, P.~Germano, {\em et~al.}, ``Jupiter rising: A decade
  of clos topologies and centralized control in google's datacenter network,''
  {\em ACM SIGCOMM computer communication review}, vol.~45, no.~4,
  pp.~183--197, 2015.

\bibitem{f10}
V.~Liu, D.~Halperin, A.~Krishnamurthy, and T.~Anderson, ``F10: A fault-tolerant
  engineered network,'' in {\em Presented as part of the 10th $\{$USENIX$\}$
  Symposium on Networked Systems Design and Implementation (NSDI)},
  pp.~399--412, 2013.

\bibitem{mdcube}
H.~Wu, G.~Lu, D.~Li, C.~Guo, and Y.~Zhang, ``Mdcube: a high performance network
  structure for modular data center interconnection,'' in {\em Proceedings of
  the 5th international conference on Emerging networking experiments and
  technologies}, pp.~25--36, ACM, 2009.

\bibitem{jellyfish}
A.~Singla, C.-Y. Hong, L.~Popa, and P.~B. Godfrey, ``Jellyfish: Networking data
  centers, randomly.,'' in {\em Proc. USENIX Symposium on Networked Systems
  Design and Implementation (NSDI)}, vol.~12, pp.~17--17, 2012.

\bibitem{kandula2009nature}
S.~Kandula, S.~Sengupta, A.~Greenberg, P.~Patel, and R.~Chaiken, ``The nature
  of data center traffic: measurements \& analysis,'' in {\em Proc. 9th ACM
  SIGCOMM conference on Internet measurement}, pp.~202--208, 2009.

\bibitem{mogul2012we}
J.~C. Mogul and L.~Popa, ``What we talk about when we talk about cloud network
  performance,'' {\em ACM SIGCOMM Computer Communication Review}, vol.~42,
  no.~5, pp.~44--48, 2012.

\bibitem{DBLP:journals/cn/ZouW0HCLXH14}
S.~Zou, X.~Wen, K.~Chen, S.~Huang, Y.~Chen, Y.~Liu, Y.~Xia, and C.~Hu,
  ``Virtualknotter: Online virtual machine shuffling for congestion resolving
  in virtualized datacenter,'' {\em Computer Networks}, vol.~67, pp.~141--153,
  2014.

\bibitem{datacenter_burstiness}
Q.~Zhang, V.~Liu, H.~Zeng, and A.~Krishnamurthy, ``High-resolution measurement
  of data center microbursts,'' in {\em Proceedings of the 2017 Internet
  Measurement Conference}, IMC '17, (New York, NY, USA), pp.~78--85, ACM, 2017.

\bibitem{proteus}
A.~Singla, A.~Singh, K.~Ramachandran, L.~Xu, and Y.~Zhang, ``Proteus: a
  topology malleable data center network,'' in {\em Proceedings of the 9th ACM
  SIGCOMM Workshop on Hot Topics in Networks}, p.~8, ACM, 2010.

\bibitem{osa}
K.~{Chen}, A.~{Singla}, A.~{Singh}, K.~{Ramachandran}, L.~{Xu}, Y.~{Zhang},
  X.~{Wen}, and Y.~{Chen}, ``Osa: An optical switching architecture for data
  center networks with unprecedented flexibility,'' {\em IEEE/ACM Transactions
  on Networking}, vol.~22, pp.~498--511, April 2014.

\bibitem{dan}
C.~Avin, K.~Mondal, and S.~Schmid, ``Demand-aware network designs of bounded
  degree,'' in {\em Proc. International Symposium on Distributed Computing
  (DISC)}, 2017.

\bibitem{venkatakrishnan2018costly}
S.~B. Venkatakrishnan, M.~Alizadeh, and P.~Viswanath, ``Costly circuits,
  submodular schedules and approximate carath{\'e}odory theorems,'' {\em
  Queueing Systems}, vol.~88, no.~3-4, pp.~311--347, 2018.

\bibitem{mordia}
G.~Porter, R.~Strong, N.~Farrington, A.~Forencich, P.~Chen-Sun, T.~Rosing,
  Y.~Fainman, G.~Papen, and A.~Vahdat, ``Integrating microsecond circuit
  switching into the data center,'' in {\em Proceedings of the ACM SIGCOMM 2013
  Conference on SIGCOMM}, SIGCOMM '13, (New York, NY, USA), pp.~447--458,
  Association for Computing Machinery, 2013.

\bibitem{perf20bmatch}
M.~Bienkowski, D.~Fuchssteiner, J.~Marcinkowski, and S.~Schmid, ``Online
  dynamic b-matching with applications to reconfigurable datacenter networks,''
  in {\em 38th International Symposium on Computer Performance, Modeling,
  Measurements and Evaluation (PERFORMANCE)}, 2014.

\bibitem{avin2019renets}
C.~Avin and S.~Schmid, ``Renets: Toward statically optimal self-adjusting
  networks,'' {\em arXiv preprint arXiv:1904.03263}, 2019.

\bibitem{schwartz2019online}
R.~Schwartz, M.~Singh, and S.~Yazdanbod, ``Online and offline greedy algorithms
  for routing with switching costs,'' {\em arXiv preprint arXiv:1905.02800},
  2019.

\bibitem{mckeown1999achieving}
N.~McKeown, A.~Mekkittikul, V.~Anantharam, and J.~Walrand, ``Achieving 100\%
  throughput in an input-queued switch,'' {\em IEEE Transactions on
  Communications}, vol.~47, no.~8, pp.~1260--1267, 1999.

\bibitem{flexspander}
M.~Y. Teh, Z.~Wu, and K.~Bergman, ``Flexspander: augmenting expander networks
  in high-performance systems with optical bandwidth steering,'' {\em IEEE/OSA
  Journal of Optical Communications and Networking}, vol.~12, no.~4,
  pp.~B44--B54, 2020.

\end{thebibliography}
}

\end{document}